\documentclass[preprint,journal]{vgtc}            % preprint (journal style)

\newcommand{\revision}[1]{\leavevmode{\textcolor[HTML]{000000}{#1}}}

\newcommand{\attention}[1]{\leavevmode{\textcolor[HTML]{000000}{#1}}}

\onlineid{2024}

\vgtccategory{Research}

\title{\sysname: Visual Exploration of Data Fact Outliers}

\author{%
  \authororcid{Yikai Li}{0009-0008-1331-5534} and
  \authororcid{Yong Wang}{0000-0002-0092-0793}
}

\authorfooter{
  \item
    Yikai Li and Yong Wang are with Nanyang Technological University.
    Yong Wang is the corresponding author.
    E-mails: liyi0084@e.ntu.edu.sg and yong-wang@ntu.edu.sg.
}

\abstract{%
  Exploratory Data Analysis (EDA) systems extract and present data facts to summarize meaningful patterns such as trends and correlations for efficient dataset exploration. However, existing approaches rarely consider outlier detection at the level of data facts, and heterogeneous facts from different analytical scopes are often aggregated in a single view, making it difficult to define meaningful metrics and effectively analyze data fact outliers.
  To fill this gap, we present \sysname, a novel visual analytics system for interactive data \underline{\textbf{F}}act \underline{\textbf{O}}utlier e\underline{\textbf{X}}ploration. 
  \sysname organizes data facts into groups with consistent analytical scopes and computes a unified outlier score that combines distribution-based and pattern-based components. Its interface comprises an \textit{Upload Panel} for data preparation and two coordinated exploration panels: the \textit{Overview Panel} employs a matrix-based visualization to enable an intuitive overview of all data facts, and the \textit{Main Panel} provides four linked views for cluster-level and fact-level analysis. We evaluated the usability and effectiveness of the system through two usage scenarios on public datasets and in-depth interviews with 12 participants. The results show that FOX enables meaningful detection, analysis, and explanation of data fact outliers.
  
}

\keywords{Data fact outlier, visual analytics, exploratory data analysis}

\teaser{
  \centering
  \includegraphics[width=\linewidth, alt={A user interface of \sysname.}]{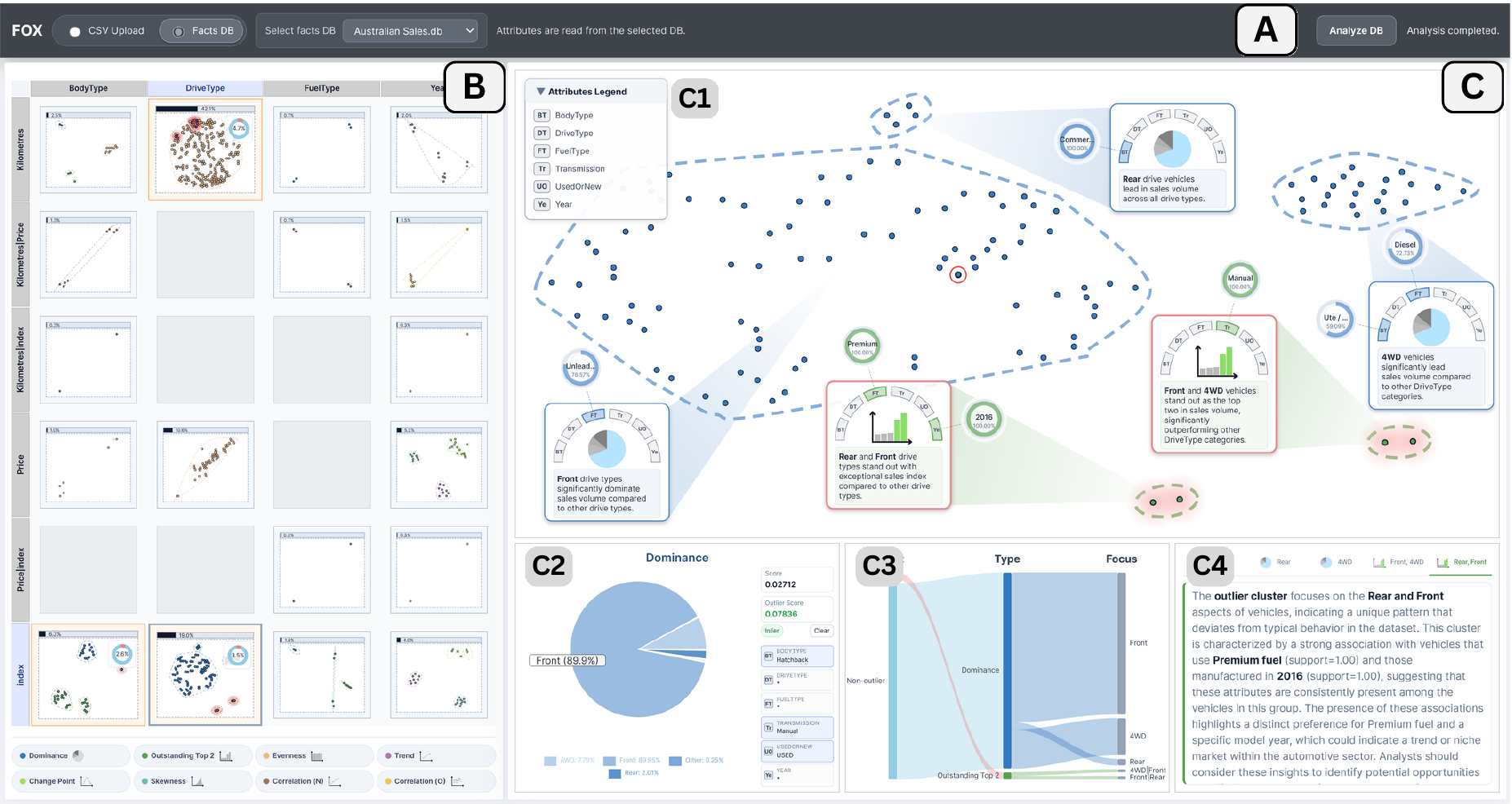}
  \caption{
    The user interface of \sysname, which is organized into two parts: preparation and exploration. The first part is supported by the Upload Panel (A), which enables dataset selection, fact extraction, and preprocessing (including grouping, similarity computation, and outlier scoring). The second part consists of the Overview Panel (B), which presents grouped facts as an interactive matrix, and the Main Panel (C), which supports detailed analysis through four views: Cluster Map (C1), Fact Detail (C2), Fact Flow (C3), and Cluster Insights (C4).
  }
  \label{fig:teaser}
}

\graphicspath{{figs/}{figures/}{pictures/}{images/}{./}} % where to search for the images

\usepackage{tabu}                      % only used for the table example
\usepackage{booktabs}                  % only used for the table example
\usepackage{lipsum}                    % used to generate placeholder text
\usepackage{mwe}                       % used to generate placeholder figures
\usepackage{ccicons}                   % package to be able to use icons from creative commons

\usepackage{mathptmx}                  % use matching math font
\usepackage{amsmath}                   % for \operatorname in math mode
\usepackage[pagebackref,bookmarks]{hyperref}
\usepackage{enumitem}

\newcommand{\sysname}{\textit{FOX}\xspace}

\begin{document}

%%%%%%%%%%%%%%%%%%%%%%%%%%%%%%%%%%%%%%%%%%%%%%%%%%%%%%%%%%%%%%%%
%%%%%%%%%%%%%%%%%%%%%% START OF THE PAPER %%%%%%%%%%%%%%%%%%%%%%
%%%%%%%%%%%%%%%%%%%%%%%%%%%%%%%%%%%%%%%%%%%%%%%%%%%%%%%%%%%%%%%%

%% The ``\maketitle'' command must be the first command after the
%% ``\begin{document}'' command. It prepares and prints the title block.
%% the only exception to this rule is the \firstsection command
\firstsection{Introduction}

\maketitle

Exploratory Data Analysis (EDA) aims to uncover patterns and derive insights from data \cite{EDA1, EDA2}.
To facilitate effective EDA, recent research has developed automated data-mining approaches~\cite{QuickInsights, Top-kInsights} and interactive visual analytics systems~\cite{InsightMap, InsightsSurvey} that extract and present \revision{\textit{data facts}~\cite{QuickInsights, InsightsSurvey}, which are interesting patterns in a dataset, such as trends and correlations, and are often represented in a tuple-based format~\cite{QuickInsights}}.
% To facilitate effective EDA, recent research has further presented automated data mining approaches~\cite{QuickInsights, Top-kInsights} and interactive visual analytics systems~\cite{InsightMap, InsightsSurvey} of \revision{\textit{data facts}, which refers to interesting data patterns of a dataset such as trends and correlations~\cite{QuickInsights, InsightsSurvey}. Such data facts are often represented in a tuple-based format~\cite{QuickInsights}}.
% Recent automated data exploration and visual analytics systems often extract and present data facts\revision{—structured representations of interesting data patterns expressed in a tuple-based format \cite{QuickInsights, InsightsSurvey}—}which summarize meaningful patterns such as trends or correlations to assist analysts in exploring datasets.

% \revision{While some visualization contexts treat data facts as simple chart captions or textual descriptions \cite{AugmentingVisualization}, we adopt a structured, schema-driven definition rooted in automated insight mining \cite{QuickInsights, Top-kInsights, Metainsight}. In this framework, a data fact is a well-defined analytical unit that binds a specific analytical scope to a characterized statistical pattern type evaluated over aggregated measures.}
% Meanwhile, outlier detection is vital in data analysis because unusual patterns may reveal valuable or unexpected insights. For instance, rare responses to medications can indicate adverse drug reactions in biology, while infrequent events may signal potential threats in network security \cite{RarePatternMining1}.

Outlier detection plays an important role in data analysis by helping analysts identify unexpected patterns that may reveal valuable insights \cite{OutlierDetection1}. However, existing approaches rarely examine outliers at the level of data facts. \attention{We define a \emph{data fact outlier} as a data fact whose pattern deviates substantially from other facts within the same analytical scope. For example, in a retail sales dataset, most facts concerning the relationship between discounts and sales exhibit a positive correlation, whereas a fact restricted to technology products shipped via second-class delivery indicates a negative correlation. Such a context-specific deviation can reveal potentially valuable behavior that would be difficult to identify through conventional analysis. Nevertheless, methods for detecting and interactively exploring these data fact outliers remain underexplored.} Although rare-pattern mining and outlier-detection methods have been extensively studied \cite{RarePatternMining1, RarePatternMining2, OutlierDetection1}, they generally focus on raw data instances, transactions, or individual patterns rather than deviations among a collection of comparable data facts. Similarly, existing visual analytics systems primarily present data facts to support exploration or storytelling \cite{Calliope, Inksight}, without explicitly identifying and supporting the investigation of data fact outliers.

Supporting such investigation presents two related challenges: ensuring meaningful comparisons among data facts and maintaining a clear visual representation. Existing visualization approaches often aggregate heterogeneous data facts into a single view to reveal global patterns and distributions \cite{InsightMap, FactExplorer}. Although this strategy is useful for obtaining an overall summary, it is problematic for outlier analysis because data facts may belong to different analytical scopes. For example, facts describing sales across countries and those describing sales across product types use different breakdown attributes and therefore cannot be meaningfully compared when computing outlier scores. Reliable fact-level outlier analysis consequently requires data facts to be organized into consistent analytical scopes, within which tailored comparison and scoring metrics can be defined. Moreover, displaying a large number of heterogeneous facts in one view can produce substantial visual clutter, making unusual patterns difficult to identify and inspect.

To address these challenges, we propose \sysname, a novel visual analytics system for data \textbf{F}act \textbf{O}utlier e\textbf{X}ploration that enables users to conduct data fact outlier analysis within each group through coordinated views, as shown in \hyperref[fig:teaser]{Fig.~\ref*{fig:teaser}}. The system automatically extracts data facts from datasets, organizes them into groups with consistent analytical scopes, computes numerical and categorical similarities, and derives outlier scores by combining distribution-based and pattern-based scores. The Upload Panel allows users to upload a target dataset and configure the settings for fact extraction. The Overview Panel presents the global distribution of facts using a matrix-based visualization and integrates interactive exploration features that allow users to select cells of interest for deeper analysis. For a selected cell, analysts can further explore the data in the Main Panel through coordinated views, where the Cluster Map view visualizes facts as a scatterplot with tailored labels and contextual descriptions, the Fact Detail view enables inspection of selected facts, the Fact Flow view provides an overview of fact distributions within the group, and the Cluster Insights view generates descriptions to summarize clusters and highlight potentially valuable insights. In summary, the main contributions of this paper are as follows:
\revision{
\begin{itemize}[leftmargin=10pt]
    \item We present \sysname, a novel visual analytics system that extracts data facts and organizes them into groups to facilitate outlier analysis. We evaluate the system through two usage scenarios and in-depth interviews with 12 participants, demonstrating its effectiveness and usability. The source code is publicly available online\footnote{\url{https://github.com/lyk6666/FOX}}.
    \item We propose a unified outlier scoring framework with tailored metrics for data fact outliers, enabling the quantification of outlier scores by combining distribution-based and pattern-based characteristics.
\end{itemize}}

\section{Related Work}
In this section, we review existing work on automatic fact mining tools, as well as methods for outlier detection and rare pattern mining.

\subsection{Automatic Fact Mining Tools}
% Over the past decade, previous EDA tools developed a structured framework and explicit definitions on the data fact extraction to mining interesting pattern in the dataset. Early systems such as Foresight \cite{Foresight} conducted data analysis through guideposts and introduce statistical descriptors, acting as an initial version of data fact; Top-k insights first propose the concept of data fact, which captures interesting observation derived from aggregation results in multiple steps to help analysts to save manual efforts through automatic extraction process; Then, QuickInsights \cite{QuickInsights} further proposes a unified formulation, more comprehensive, of data fact, using a 5-tuple, and designs a systematic mining framework according to it; MetaInsight \cite{Metainsight} introduce the idea of homogeneous data pattern to group data facts, and propose a novel scoring framework for data facts. 
Over the past decade, prior EDA research has developed structured frameworks and explicit definitions for extracting data facts and mining interesting patterns from datasets. Demiralp et al. \cite{Foresight} supported rapid data exploration through guideposts and statistical descriptors, serving as an early precursor to the data fact concept. Tang et al. \cite{Top-kInsights} formally introduced the concept of data facts, capturing interesting observations derived from aggregation results to reduce analysts’ manual effort during exploration. Ding et al. \cite{QuickInsights} established a more comprehensive and unified formulation of data facts using a 5-tuple representation, together with a systematic mining framework based on this formulation. Ma et al. \cite{Metainsight} further introduced the notion of homogeneous data patterns for grouping data facts and developed a scoring framework.

Apart from theoretical design, a variety of visual fact mining and analytics systems have been proposed to leverage visual representations for improving human understanding and supporting insight discovery beyond traditional analysis methods \cite{InsightsSurvey}. 
% We examined a range of existing visual analytics tools as well as recent surveys. 
Prior work can be broadly grouped into data storytelling systems, visualization recommendation approaches, and LLM-assisted tools that enable mixed-initiative data analysis. Data storytelling systems aim to extract and structure insights into coherent narratives, such as Calliope \cite{Calliope}, which explores tabular data to construct data-driven stories, CoInsight \cite{CoInsight}, which derives and relates insights from hierarchical tables, FactExplorer \cite{FactExplorer}, which exposes the space of possible facts to support exploration, and DataShot \cite{DataShot}, which automates the generation of concise analytical summaries. Visualization recommendation approaches focus on mapping data to effective visual representations, including Data2Vis \cite{Data2Vis}, which generates visualizations from high-level specifications, DashBot \cite{Dashbot}, which learns to construct analytical dashboards, and DeepEye \cite{DeepEye}, which automatically selects data transformations and visualization types. 
More recently, LLM-assisted tools have supported interactive and mixed-initiative analysis \cite{Inksight,InsightLens,InsightPilot}. For example, InkSight \cite{Inksight} documents analytical findings, InsightLens \cite{InsightLens} helps users navigate complex conversational contexts, and InsightPilot \cite{InsightPilot} guides analysis workflows and insight generation.
Our system organizes data facts into groups and focuses on detecting and analyzing data fact outliers, with LLM-generated summaries to aid interpretation.
 
\subsection{Outlier Detection and Rare Pattern Mining}
Pattern mining was introduced by Agrawal et al. \cite{RulesMining} to discover frequent patterns in transactional data and was later extended to identify rare but meaningful patterns. Rare pattern mining methods are generally Apriori- or FP-Growth–based. MS-Apriori \cite{RarePatternMining1, MSApriori} introduced multiple minimum support thresholds, followed by extensions such as ARIMA \cite{ARIMA} and HURI \cite{HURI}. However, Apriori-based methods scale poorly because retaining rare items causes rapid candidate growth \cite{RarePatternMining1, RarePatternMining2}. FP-Growth–based methods, including MCCFP \cite{MCCFP}, MSFP \cite{MSFP}, and MCRP \cite{MCRP}, avoid explicit candidate generation. Nevertheless, most studies focus on rare association rules in static data \cite{RarePatternMining2}, while sequential rare pattern mining remains less explored due to temporal dependencies and exponential candidate growth \cite{RarePatternMining1, RarePatternMining2}.

Outlier detection identifies rare instances that deviate substantially from the majority \cite{OutlierDetection1}. Depending on label availability, methods are classified as supervised, semi-supervised, or unsupervised, with unsupervised approaches most widely studied because labeled anomalies are scarce. Existing techniques are broadly proximity- or projection-based. Proximity-based methods use neighborhood relationships: LOF \cite{LOF} measures local-density deviation, while COF \cite{COF} and LOCI \cite{LOCI} consider connectivity and multi-scale density, respectively. However, their reliance on nearest-neighbor searches can be computationally costly. Projection-based methods improve scalability through dimensionality reduction; for example, PINN \cite{PINN} uses random projections to approximate k-nearest neighbors efficiently.

% Outlier detection focuses on identifying data instances that significantly deviate from the majority and are typically characterized by feature-level deviation and rarity within the dataset \cite{OutlierDetection1}. Based on label availability, outlier detection methods are classified into supervised, semi-supervised, and unsupervised approaches, with unsupervised methods being the most widely studied due to the limited availability of labeled anomalies. Broadly, existing techniques can be divided into proximity-based and projection-based approaches. Proximity-based methods evaluate outlierness using neighborhood relationships, where the Local Outlier Factor (LOF) \cite{LOF} introduced the notion of local density deviation and inspired extensions such as COF \cite{COF}, which considers connectivity, and LOCI \cite{LOCI}, which analyzes multi-scale local density. Although effective, these methods often incur high computational costs due to nearest-neighbor searches. To improve scalability, projection-based approaches reduce data dimensionality while preserving neighborhood structure; a notable example is PINN \cite{PINN}, which applies random projection to efficiently approximate k-nearest neighbors for scalable outlier detection. 

To the best of our knowledge, our system is among the first systems to explicitly support outlier analysis at the level of data facts and proposes a new metric to identify such outliers by combining distribution-based and pattern-based features, leveraging both LOF and information-theoretic measures.

\begin{figure*}[t]
  \centering
  \includegraphics[width=1\textwidth]{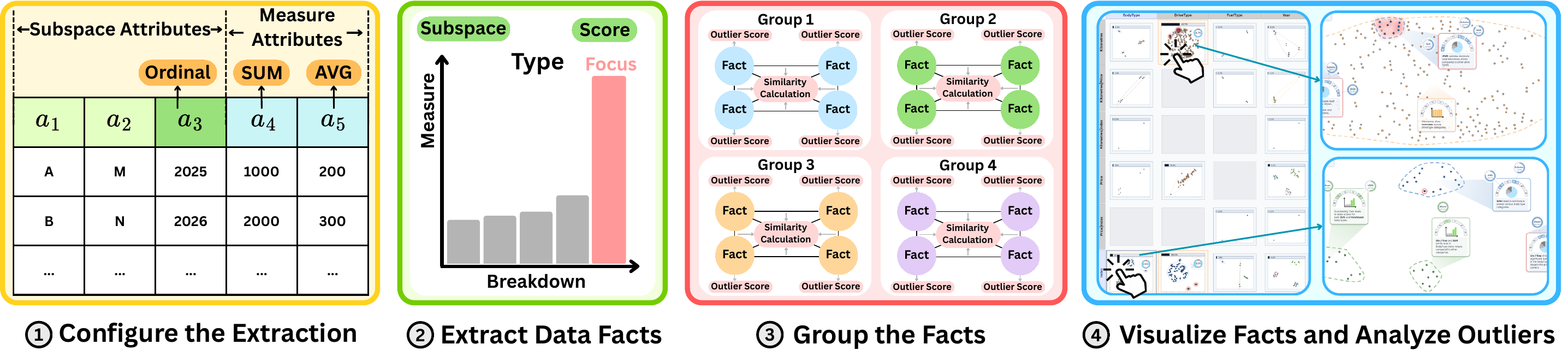}
  \caption{The pipeline consists of four stages: (1) dataset upload and configuration for subspace and measure attributes, (2) automatic fact extraction, (3) fact grouping with similarity computation and outlier scoring, and (4) an interactive interface for visualizing data facts and analyzing fact outliers through cluster exploration, fact inspection, and descriptive summaries.}
  \label{fig:framework}
\end{figure*}

\section{System Design}
In this section, we discuss the design requirements derived from prior work and present the framework of our system.

\subsection{Design Requirements} 
Our system aims to support effective dataset exploration by enabling users to conduct structured outlier analysis in an efficient and intuitive manner. Based on this goal, we derive six key design requirements from prior work that guide the development of our system, following a workflow from data processing to interactive exploration.

\textbf{R1: Automatic Fact Extraction.}
\revision{
To reduce manual effort and enable systematic data exploration, automatic data fact extraction has been widely recognized as an established practice in visual analytics research \cite{Calliope, QuickInsights, Top-kInsights}. However, prior systems often rely on prioritization strategies such as best-first search \cite{QuickInsights} to identify top-ranked facts, which may overlook facts that are less prominent under the criteria but still valuable as outliers relative to other facts. Therefore, \sysname should adopt an extraction strategy that more comprehensively covers diverse subspaces, providing a broader fact population for downstream grouping and outlier scoring.
}

% While automated fact mining is an established best practice in visual analytics to reduce manual effort \cite{Calliope, FactExplorer}, it serves as a strict functional prerequisite in FOX. Reliably detecting data fact outliers requires comprehensive fact extraction across diverse subspaces \cite{QuickInsights, Top-kInsights}. Since manual exploration introduces selection bias and omits crucial patterns, automated extraction is essential to construct the data fact populations needed for our downstream grouping and outlier scoring frameworks.
% Datasets typically comprise numerous subspaces, making it essential to efficiently identify and extract meaningful data facts while avoiding redundant computations on uninformative subspaces or insignificant results \cite{QuickInsights, Top-kInsights}. To better support subsequent fact-outlier analysis, an efficient and well-structured fact-mining mechanism is required to automate this process.

\textbf{R2: Structured Fact Grouping for Consistent Analysis.}
Presenting all data facts within a single unified view provides an intuitive overview of their global distribution and patterns \cite{FactExplorer, InsightMap}. However, effective outlier analysis requires facts to be grouped according to consistent breakdown and measure attributes \cite{QuickInsights}. Such structured grouping enables the design of outlier metrics tailored to specific analytical scopes. By ensuring that all facts within a group share the same breakdown--measure scope, pattern interpretation and comparison remain coherent and meaningful.

\textbf{R3: Tailored Metrics for Fact-Level Outlier Scoring.}
Extensive research has been conducted on outlier and anomaly detection \cite{OutlierDetection1, AnomalyDetection}. However, existing metrics are not directly applicable to the analysis of data facts. Prior work such as MetaInsight \cite{Metainsight} introduced the idea of grouping data facts into commonness and exception categories at a global level, but did not propose dedicated metrics to quantify fact-level outlierness, instead relying on simple overlap-based similarity measures. Therefore, there is a need to design system-specific and context-aware metrics that can effectively quantify outlier scores for individual data facts and support subsequent analysis.

\revision{\textbf{R4: Global Overview and Group Identification.} Analysts need to efficiently monitor the global landscape of data facts spanning multiple heterogeneous breakdown and measure attributes \cite{CoInsight}. The system should support a rapid visual scan across distinct fact groups, enabling users to immediately identify which analytical scopes contain anomalous behaviors or high proportions of outliers without initial exposure to the complexity of individual data facts.}

\revision{\textbf{R5: Intra-Group Exploration with Drill-Down.} Once an interesting analytical group is isolated, users require capabilities to progressively explore its local structure \cite{FactExplorer, InsightMap}. Analysts need to examine detailed fact distributions, inspect individual subspaces, and interactively trace how data patterns correlate across different dimensions such as outlier status, pattern types, and focuses.}

\revision{\textbf{R6: Cognitive Load Reduction.} Evaluating large clusters of data facts can be time-consuming and cognitively overwhelming, especially for users without deep domain expertise. Instead of forcing analysts to interpret abstract patterns manually, the system should automatically turn complex cluster characteristics into concise, natural-language summaries that highlight key analytical takeaways \cite{InsightPilot, InsightLens}.}

\subsection{Framework}
As illustrated in \hyperref[fig:framework]{Fig.~\ref*{fig:framework}}, the framework consists of four main stages. First, the analyst uploads a dataset and configures the analysis by specifying subspace attributes, which also serve as breakdown attributes, and measure attributes. Ordinal attributes can be selected within the subspace attributes, and an aggregation method is assigned to each measure attribute to prepare the system for fact extraction.
% \revision{FOX features an Automatic Fact Extraction framework (\textbf{R1}). While prior insight mining systems typically use best-first prioritization \cite{QuickInsights}, our framework systematically traverses the attribute space via depth-first search, leveraging an impact-based pruning strategy to early-terminate branches and maintain computational efficiency, while directly adopting established methodologies from previous literature for type extraction \cite{QuickInsights, Top-kInsights}.}
The system then automatically extracts data facts that satisfy the predefined configurations, generating a file containing all extracted facts (\textbf{R1}).

Next, the analyst selects the extracted fact file for further processing. The system organizes the facts into groups (\textbf{R2}), computes pairwise similarity within each group, and calculates an outlier score for each fact using our proposed metrics (\textbf{R3}), thereby identifying outlier facts and preparing the data for visualization and exploration. After processing, the analyst can interactively explore the facts and analyze fact outliers through the system interface (\textbf{R4}). By selecting groups of interest (\textbf{R5}), users can explore detailed information through coordinated views. These views support the inspection of data facts and their relationships across contexts, while LLM-assisted summaries and labels are generated through a designed prompting strategy (\textbf{R6}).

\section{Data Fact Processing}
This section describes the data fact processing pipeline, which entails extracting insightful data facts from the dataset, grouping them according to predefined rules, and calculating their outlier scores.

\subsection{Fact Extraction}
\label{sec:facts_extraction}
Data facts capture meaningful patterns hidden in data and serve as fundamental building blocks for data exploration. Following prior work \cite{QuickInsights, Metainsight, Top-kInsights}, we adopt a formalized representation of data facts to systematically extract valuable insights that satisfy predefined requirements. 
In our formulation, each data fact is defined by six core elements and represented as a 6-tuple:

\[ \{ subspace, breakdown, measure, type, score, focus \}. \]
\revision{Consider a hypothetical global retail dataset comprising annual sales records from different countries and three attributes: \textit{Country}, \textit{Year}, and \textit{Sales}. A data fact derived from this dataset may characterize the distribution of total sales across countries in a given year, identifying the dominant country and quantifying the interestingness of the observed pattern. For example, such a fact can be represented as $\{\{Country = *, Year = 2025\},\; Country,\; Sales,\; Dominance,\; 0.5,\; \text{\textit{Country A}} \}$.}

\begin{figure}[t]
  \centering
  \includegraphics[width=0.95\linewidth]{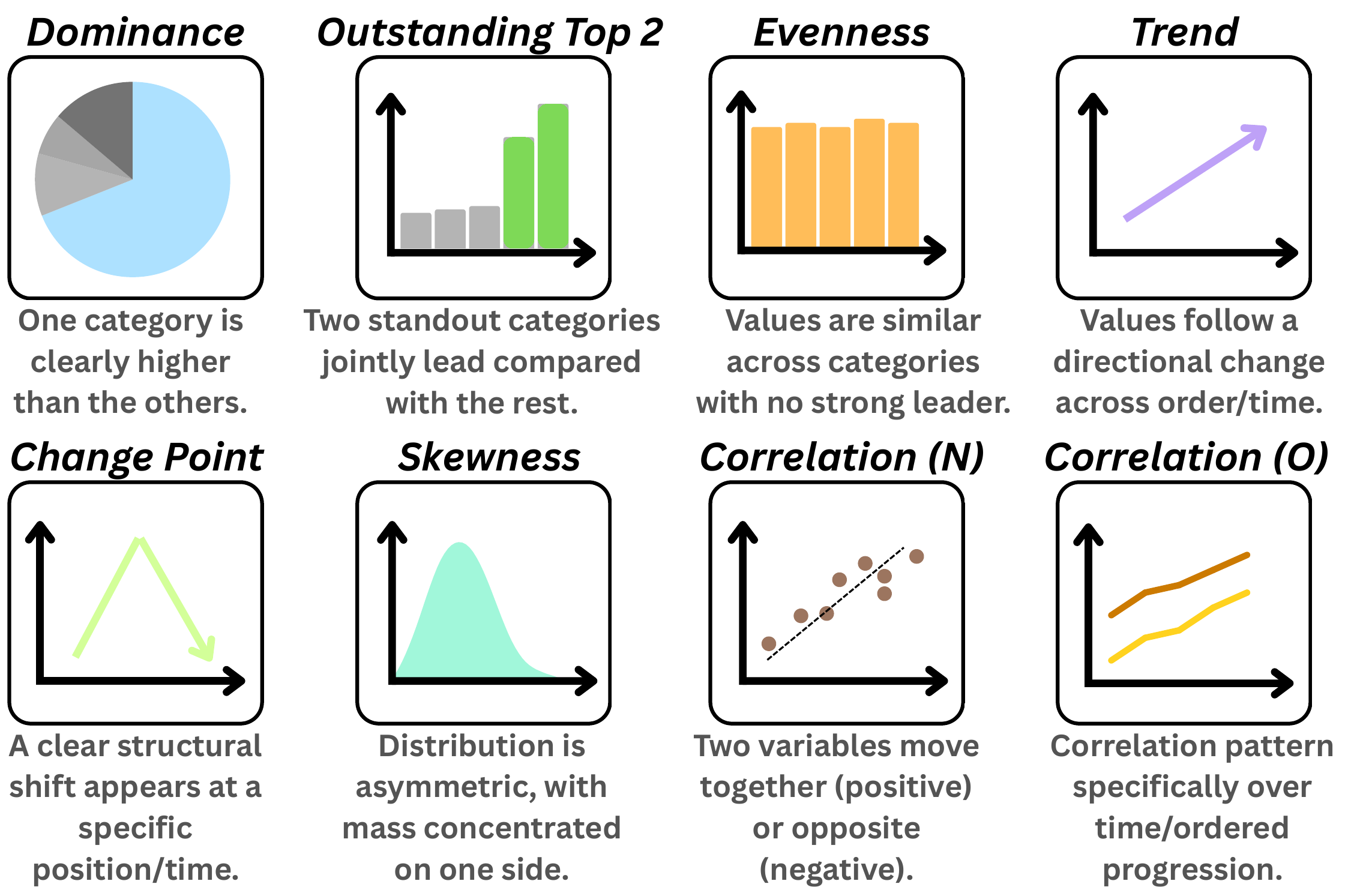}
  \caption{Our system supports eight types of data facts: \textit{Dominance}, \textit{Outstanding Top 2}, \textit{Evenness}, \textit{Trend}, \textit{Change Point}, \textit{Skewness}, \textit{Correlation (N)}, and \textit{Correlation (O)}, each accompanied by a brief description. Here, \textit{Correlation (N)} refers to correlations over nominal attributes, while \textit{Correlation (O)} refers to correlations over ordinal attributes.}
  \label{fig:TypeDescription}
\end{figure}

\textbf{Subspace} defines the analytical scope of a data fact through attribute filters over nominal and ordinal attributes, which we collectively refer to as subspace attributes. It is represented as a fixed-length tuple of size $n$, where the $k$-th attribute $a_k$ takes value $v_k \in V_k \cup \{*\}$, and $V_k$ denotes the set of all distinct values of $a_k$. The wildcard $*$ indicates that no filter is applied to the corresponding 
attribute, and the number of specified attribute values determines the subspace degree. 
Formally, a subspace is defined as:
\[
\{\{a_1 = v_1\}, \{a_2 = v_2\}, \dots, \{a_n = v_n\}\}, 
\quad v_k \in V_k \cup \{*\}.
\]
\revision{In our example, the subspace $\{Country = *, Year = 2025\}$ defines the fact’s analytical context as records in the year 2025.}
\textbf{Breakdown} performs a group-by operation on a subspace using an attribute without an applied filter, partitioning the data for aggregation.
\textbf{Measure} refers to one or two numerical attributes aggregated using functions such as sum, average, minimum, or maximum. 
\textbf{Type} describes the pattern category of a data fact. Following prior work \cite{Calliope-Net, Top-kInsights, QuickInsights}, we adopt eight types in our system, as illustrated in \hyperref[fig:TypeDescription]{Fig.~\ref*{fig:TypeDescription}}, where each type is encoded with a distinctive color consistent with the system design.
\revision{Selecting \textit{Country} as the breakdown and \textit{Sales} as the measure implies that the fact evaluates the distribution of sales volumes across countries within the subspace.}
\textbf{Score} quantifies the interestingness of a fact using impact and significance. Impact is computed using a count aggregation that satisfies the anti-monotonicity property \cite{Anti-Monotonic, QuickInsights},
while significance follows prior formulations \cite{Top-kInsights, QuickInsights}. 
The final score is defined as:
\begin{equation}
score_i = impact_i \times significance_i.
\end{equation}
\textbf{Focus} highlights the key element of a fact that requires attention, 
\revision{such as a \textit{Dominance} type with a focus on \textit{Country A}, which indicates that \textit{Country A} leads sales within this specific context with a score of 0.5.}

\attention{Based on the above representation, \sysname automatically extracts data facts by enumerating candidate subspaces within the user-configured attribute space. Unlike prior systems that prioritize fact extraction tasks using impact-based best-first search \cite{QuickInsights}, \sysname employs depth-first search to systematically traverse the attribute space and retain all facts that satisfy the predefined extraction criteria (\textbf{R1}). To improve computational efficiency, the system applies impact-based pruning \cite{QuickInsights} to terminate branches that cannot produce eligible facts, while the fact types and their associated scores are computed using established methods from prior work \cite{QuickInsights, Top-kInsights}.}
% \(
% impact_i = \frac{\operatorname{COUNT}(subspace_i)}{\operatorname{COUNT}(\{*\})}
% \)

% such as the dominant category in a Dominance pattern or the direction of change in a Trend pattern.

% To further clarify the definition of a data fact, we provide the following example. Consider the data fact
% $\{\{Country = *, Year = 2025\},\; Country,\; Sales,\; Dominance,\; 0.5,\; A \}$. This fact indicates that in 2025, Country A dominates the sales among all countries, with a score of 0.5.

\subsection{Fact Grouping}
% Since data facts are generated under different breakdown–measure combinations, analyzing them within a single global scope is inappropriate. For example, comparing facts about sales across years with facts about discounts across countries is not meaningful, as they describe patterns in fundamentally different measure-breakdown scopes. Performing outlier detection across such heterogeneous scopes may therefore lead to uninterpretable results. In addition, measuring similarity between data facts becomes difficult when they originate from different scopes, as cross-scope comparisons are hard to define in an explainable manner.

% One possible approach is to perform local outlier analysis within each smaller cluster in the global view \cite{OutlierDetection1, LOF}. However, further subdividing facts into smaller clusters may reduce sensitivity in detecting meaningful deviations. To ensure meaningful comparisons and more reliable similarity measurements, we restrict outlier analysis to facts sharing the same breakdown and measure. Accordingly, we design a grouping mechanism that partitions facts by their breakdown and measure (\textbf{R2}). Specifically, two data facts belong to the same group if they share identical breakdown and measure attributes, i.e., the same breakdown–measure scope.

Since data facts are generated under distinct breakdown–measure combinations, analyzing them within a single global scope is inappropriate. For example, comparing facts about sales across years with facts about discounts across countries is conceptually inconsistent because they are defined over different analytical scopes, meaning that they involve different breakdown or measure attributes. Performing outlier detection or similarity measurements across different scopes yields uninterpretable results because cross-scope comparisons lack explainability.

While local outlier analysis within global clusters is a potential alternative \cite{OutlierDetection1, LOF}, further subdividing facts can reduce the system's sensitivity to meaningful deviations. To ensure valid and effective comparisons, we restrict analysis to facts sharing identical breakdown and measure attributes. Accordingly, we introduce a grouping mechanism that partitions data facts by identical breakdown and measure attributes, restricting each analysis to a specific scope (\textbf{R2}).

\subsection{Similarity}
After grouping data facts by breakdown and measure, we compare facts within each group to support outlier detection. Since grouped facts share the same analytical scope, they can be represented consistently and compared meaningfully. Our similarity measure combines numerical similarity between aggregated value distributions with categorical similarity based on fact type and focus, as illustrated in \hyperref[fig:scoreMechanism]{Fig.~\ref*{fig:scoreMechanism}}.

Each data fact aggregates records according to the breakdown attribute and computes aggregated values based on the measure attribute, producing a set of aggregated results. These results can be represented as a vector in which the keys correspond to breakdown attribute values and the entries represent the aggregated measure values. Since facts within the same group share the same breakdown attribute, their vectors have identical dimensions and key sets. Therefore, our grouping mechanism provides a consistent representation that enables meaningful similarity comparison.

Although different subspaces may contain varying numbers of records, resulting in different aggregated magnitudes, our analysis focuses on the distributions of values rather than their absolute scales. For this reason, we adopt cosine similarity to measure the similarity between data facts \cite{Cosine-Similarity1, Cosine-Similarity2}.
Let the measure attribute be $m$, let $v_k^i$ denote 
the $i$-th distinct value in $V_k$, and let $m^i$ denote the aggregated measure 
value corresponding to $v_k^i$. The vector representation of a fact can therefore 
be written as:
\[
\{\{v_k^1 : m^1\}, \{v_k^2 : m^2\}, \dots, \{v_k^{|V_k|} : m^{|V_k|}\}\}.
\]
Given two facts $f_a$ and $f_b$ in the same group, their numerical similarity is 
measured using cosine similarity \cite{Cosine-Similarity1, Cosine-Similarity2}:

\begin{equation}
    sim_{cos}(f_a, f_b) =
    \frac{\sum_{i=1}^{|V_k|} m_a^i m_b^i}
    {\sqrt{\sum_{i=1}^{|V_k|} (m_a^i)^2}
    \sqrt{\sum_{i=1}^{|V_k|} (m_b^i)^2}}.
\end{equation}
This metric captures the similarity between the value distributions of the two facts while being invariant to their absolute magnitudes.

However, relying solely on cosine similarity is insufficient in the context of 
data facts. In practice, we observe several limitations. For example, when the 
number of breakdown categories $|V_k|$ is large, cosine similarity may dilute 
the influence of the main characteristic of a fact, reducing its discriminative 
power. In addition, small differences in value distributions may lead to 
substantially different interpretations. For instance, a proportion of $49\%$ 
versus $51\%$ may produce similar cosine similarity values, yet the latter 
indicates a \textit{Dominance} type.

To better capture the categorical characteristics of data facts, we additionally incorporate the type and focus into the similarity calculation. They explicitly describe the type category and the key element of a fact, providing complementary information beyond numerical 
distributions. We therefore adopt a simple overlap metric \cite{CategoricalSimilarity} for these categorical attributes, where the similarity equals $1$ if two values are identical and $0$ otherwise, denoted as $sim_{type}$ and $sim_{focus}$, respectively.
The final similarity between two data facts is 
defined as a weighted combination of numerical and categorical similarities:
\begin{equation}
    sim(f_a, f_b) = \alpha_1 \cdot sim_{cos} + \alpha_2 \cdot sim_{type} + \alpha_3 \cdot sim_{focus},
\end{equation}
where $\alpha_1 + \alpha_2 + \alpha_3 = 1$, and the weights control the relative importance of 
distribution similarity and categorical characteristics.

\subsection{Outlier Score}
After computing similarity, we estimate the outlier score of each data fact within its group. A straightforward approach is to apply LOF \cite{LOF, OutlierDetection1} to the similarity-based distance matrix to detect distribution anomalies. However, LOF alone is not sufficient in practice. Different groups may exhibit distinct distribution characteristics, and a single parameter setting cannot reliably capture all meaningful deviations. As a result, facts that are statistically close to others but exhibit distinctive pattern-level characteristics may remain undetected.

% Previous studies on data fact analysis \cite{Metainsight} report that such deviations often arise from uncommon fact characteristics. Two common situations can occur. First, certain fact types may dominate within a group while other types appear rarely. Second, within the same fact type, some focuses may occur frequently while others appear only a few times. For example, sales across countries may typically show an even distribution, while only a few facts reveal that a particular country dominates the sales. Similarly, within the dominance type, most facts may indicate that one country dominates the market, while only a few indicate dominance by another country. These rare characteristics may correspond to valuable insights. However, previous work relies on simple overlap based metrics \cite{Metainsight}, which cannot adequately capture their statistical properties.

Previous studies note that deviations among data facts often stem from uncommon characteristics \cite{Metainsight}. This typically manifests in two ways: either certain fact types occur rarely within a group dominated by others, or specific values appear infrequently within the same fact type. For example, while most facts about sales across countries show an even distribution, only a few may reveal the dominance of a single country. Similarly, within that \textit{Dominance} type, a specific country might dominate rarely compared to others. While these rare characteristics often harbor valuable insights, existing simple overlap-based metrics \cite{Metainsight} fail to adequately capture their underlying statistical properties.

To better quantify such deviations, we introduce a pattern-based outlier metric inspired by information theory \cite{InformationTheory, InformationEntropy}. The metric evaluates the rarity of pattern elements by considering two distributions: the distribution of fact types within a group and the distribution of fact focuses within each fact type. Let $p_t$ denote the probability of observing a pattern element $t$, which corresponds to either the probability of a fact type within a group or the probability of a fact focus within a specific fact type. The pattern-based outlier score is defined as:
\begin{equation}
S_{Pat}(t) = (1 - p_t)(1 - E), \quad
E = \frac{-\sum_i p_i \log p_i}{\log K}.
\label{eq:patternscore}
\end{equation}
Here, $K$ denotes the number of possible element values, corresponding to either fact types or focuses. The entropy term captures the overall skewness of the distribution, while $(1 - p_t)$ reflects the outlierness of a specific element. Together, these two factors characterize the degree to which an element deviates from the prevailing patterns in the distribution. Consequently, elements that occur infrequently within a highly skewed distribution receive higher scores, indicating stronger outlierness in terms of pattern characteristics.

\begin{figure}[t]
  \centering
  \includegraphics[width=0.95\linewidth]{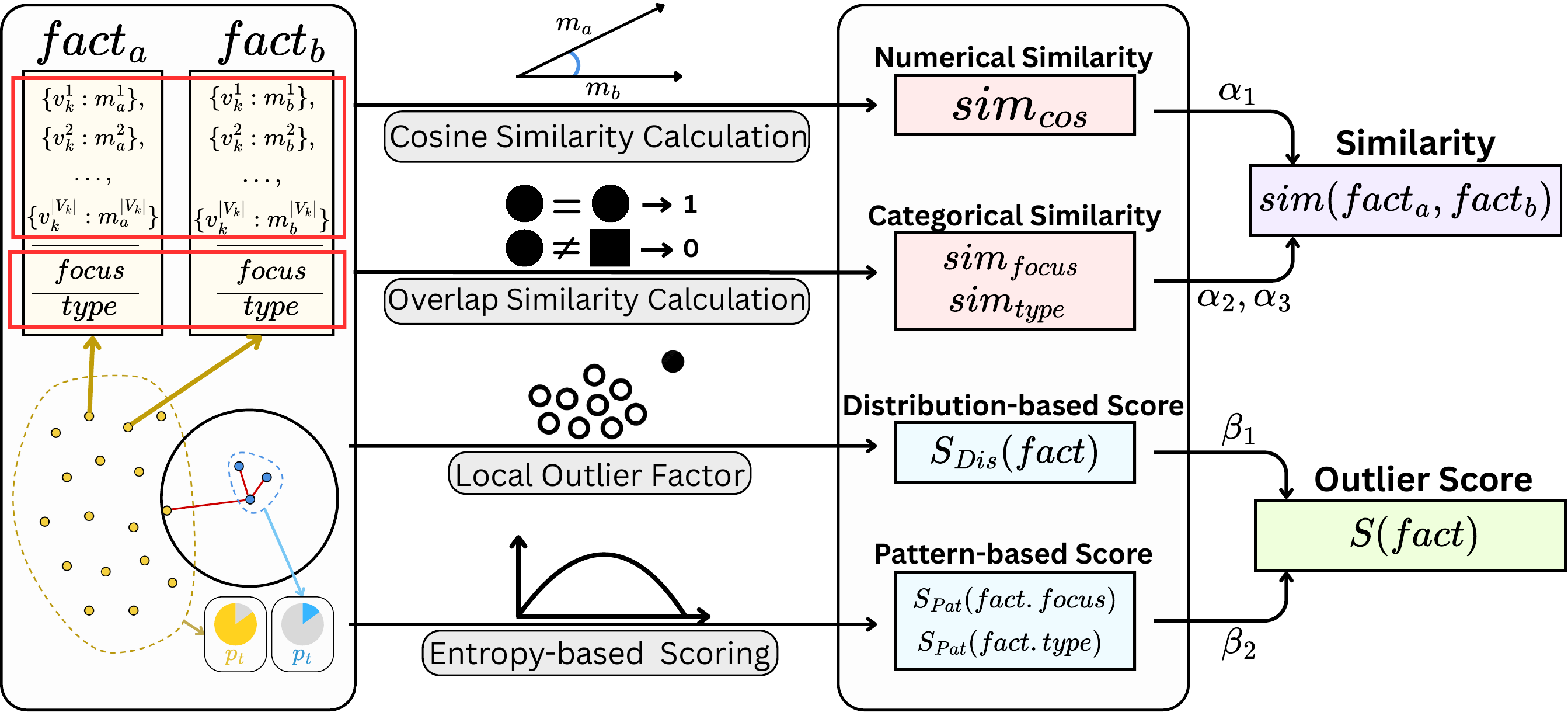}
  \caption{Computation of similarity and outlier scores. Numerical and categorical similarities are combined to compute fact similarity, while the final outlier score integrates a distribution-based score using LOF and a pattern-based score derived from entropy.}
  \label{fig:scoreMechanism}
\end{figure}

\begin{figure}[t]
  \centering
  \includegraphics[width=1\linewidth]{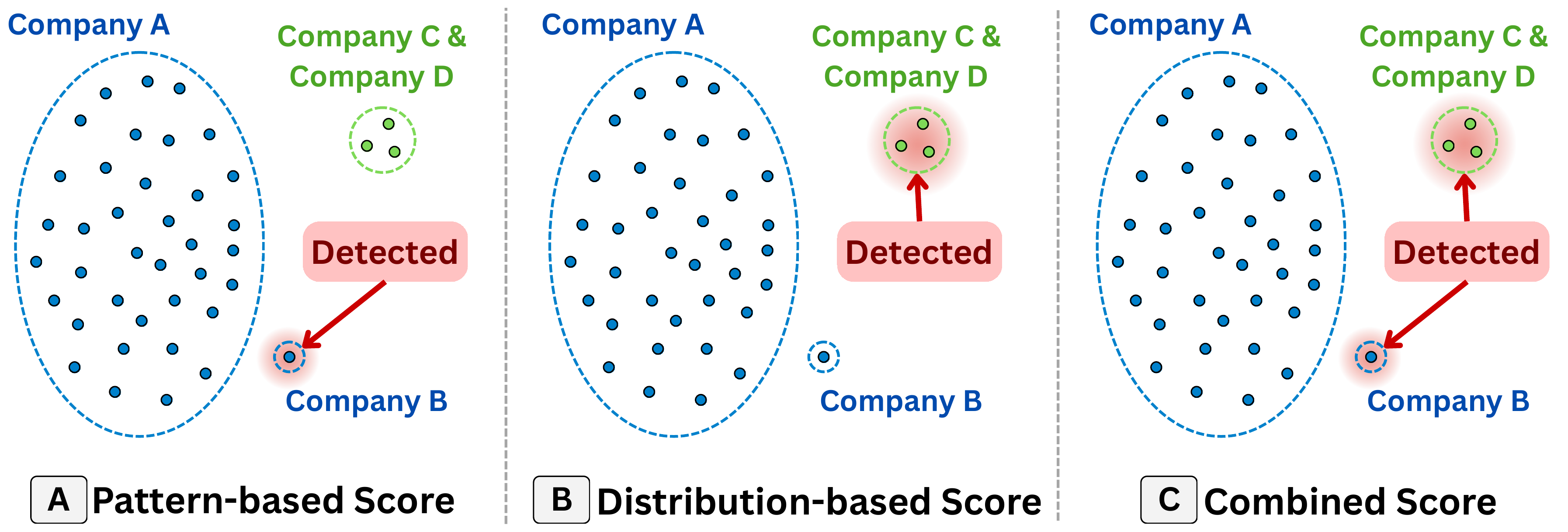}
  \caption{An example illustrating the complementary roles of the two outlier scoring components. Three clusters are constructed: a majority cluster (\textit{Dominance}, \textit{Company A}), a cluster with one fact (\textit{Dominance}, \textit{Company B}), and a cluster with three facts (\textit{Outstanding Top 2}, \textit{Company C \& Company D}). (A) The pattern-based score detects the one-fact cluster but misses the three-fact cluster. (B) The distribution-based score detects the three-fact cluster but misses the one-fact cluster. (C) The combined score detects both.}
  \label{fig:scoreevaluation}
\end{figure}

% To obtain the final outlier score for each data fact, we combine a distribution-based score with a pattern-based score. The distribution-based score is computed via LOF in the original space, prior to dimensionality reduction, to preserve accurate distance relationships. In contrast, the pattern-based score captures pattern-level deviations in fact types and focuses. These two perspectives are complementary: the distribution score identifies facts with anomalous value distributions, while the pattern score highlights rare pattern-level characteristics within the group. To integrate them, we normalize the LOF score and denote it as $S_{Dis}$ \cite{OutlierNormalization}. Since outlierness may arise from either fact type or focus, we take the maximum of the two pattern-based scores as the pattern signal. 

To obtain the final outlier score for each data fact, we combine a distribution-based score with a pattern-based score. The distribution-based score is computed via LOF using the combined similarity-based distance matrix prior to dimensionality reduction. It captures local deviations in neighborhood structure based on numerical distributions, fact types, and focuses. The pattern-based score separately quantifies the group-level rarity of fact types and focuses. They capture different aspects of outlierness: the distribution-based score evaluates pairwise neighborhood relationships, whereas the pattern-based score characterizes categorical prevalence within the group. To integrate them, we normalize the LOF score and denote it as $S_{Dis}$~\cite{OutlierNormalization}. Since outlierness may arise from either fact type or focus, we take the maximum of the two pattern-based scores as the pattern signal.
The final outlier score of a data fact $f$ is defined as:
\begin{equation}
S(f) =
\beta_1 \cdot S_{Dis}(f)
+
\beta_2 \cdot \max\left(S_{Pat}(f.type), S_{Pat}(f.focus)\right),
\end{equation}
where $\beta_1$ and $\beta_2$ control the relative contributions of the 
distributional outlier score and the pattern-based outlier score, with 
$\beta_1 + \beta_2 = 1$. The resulting value represents the overall outlier 
score of a data fact. A fact is identified as an outlier if its score exceeds 
a predefined threshold. \hyperref[fig:scoreMechanism]{Fig.~\ref*{fig:scoreMechanism}} provides an intuitive illustration of how similarity and outlier scores are computed (\textbf{R3}).

Since the proposed outlier score targets fact-level anomalies, a concept not explicitly addressed by existing metrics, there is no direct baseline available for quantitative comparison. Therefore, we illustrate the behavior and practical utility of the proposed metric through a component analysis and two usage scenarios. These evaluations demonstrate how its components contribute to detecting different types of data fact outliers.
To illustrate the complementary roles of the two components, we construct a synthetic example consisting of three clusters of data facts. The first cluster corresponds to facts with type \textit{Dominance} and focus \textit{Company A}. These facts form the majority of data facts in the group. The second cluster contains a single fact with type \textit{Dominance} and focus \textit{Company B}, representing a rare focus within the same fact type. The third cluster contains three facts with type \textit{Outstanding Top 2} and focus \textit{Company C \& Company D}, representing a different pattern.

As shown in \hyperref[fig:scoreevaluation]{Fig.~\ref*{fig:scoreevaluation}A}, when only the pattern-based outlier score is applied, the fact associated with \textit{Company B} is detected as an outlier because its focus appears only once among facts of the same type. According to the pattern-based scoring formula in \hyperref[eq:patternscore]{Eq.~\ref{eq:patternscore}}, this rare occurrence results in a very small $p_t$, producing a high pattern-based outlier score. In contrast, the cluster involving \textit{Company C \& Company D} is not identified because this pattern occurs multiple times within the group, leading to a larger $p_t$ and therefore a lower pattern-based score that does not exceed the detection threshold. As shown in \hyperref[fig:scoreevaluation]{Fig.~\ref*{fig:scoreevaluation}B}, when only the distribution-based outlier score is applied, the cluster corresponding to \textit{Company C \& Company D} is detected because these facts are positioned farther from the majority cluster in the similarity space, resulting in a high LOF score. In contrast, the fact associated with \textit{Company B} is not detected because it lies close to the majority cluster centered around \textit{Company A} and therefore does not exhibit strong distributional deviation.
When the combined score is applied (\hyperref[fig:scoreevaluation]{Fig.~\ref*{fig:scoreevaluation}C}), both the \textit{Company B} fact and the \textit{Company C \& Company D} cluster are successfully detected. This result demonstrates that the pattern-based and distribution-based components capture complementary aspects of data fact outliers.

In addition, when applied to the datasets used in \hyperref[sec:scenario]{Sec.~\ref*{sec:scenario}}, the proposed outlier score produced results broadly consistent with our qualitative observations. These examples provide preliminary evidence that the metric can help analysts identify potentially meaningful data fact outliers during exploration.

\begin{figure*}[t]
  \centering
  \includegraphics[width=1\textwidth]{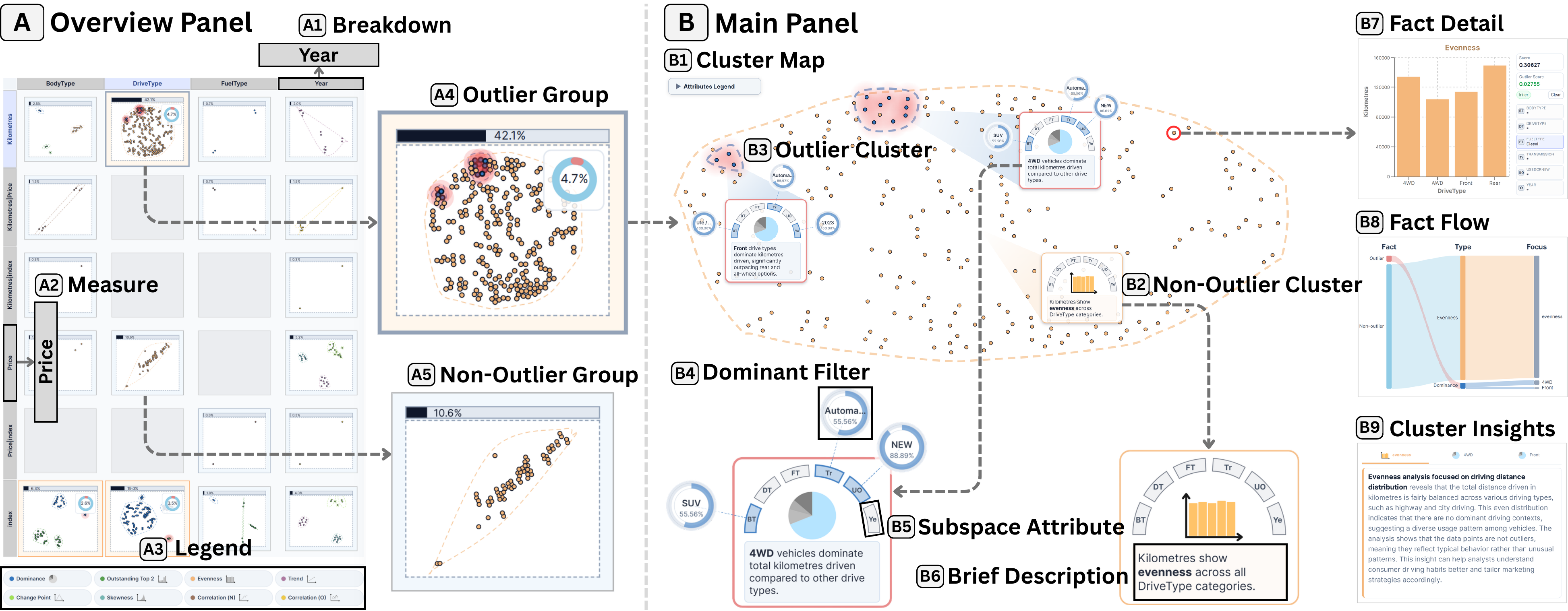}
  \caption{The user interface of FOX and its visual design. This figure presents the Overview Panel (A) and the Main Panel (B) of the system interface, illustrating the key design components and visual encodings used in the system.}
  \label{fig:VisualExplanation}
\end{figure*}

\section{Interface and Visual Design}
% \section{Visual Analytics Interface}
Our system interface is organized into two parts: preparation and exploration. The first part is supported by the Upload Panel, while the second part consists of the Overview Panel and the Main Panel.

\textbf{Upload Panel.}
% The Upload Panel consists of two stages. In the first stage, users upload a dataset and specify the analysis configuration, including the subspace and measure attributes. Ordinal attributes can be selected within the subspace attributes to represent sequential variables such as rankings or years, while aggregation methods can be assigned to each measure attribute. 
% % After the configuration is confirmed, the system automatically extracts data facts from the dataset according to predefined settings and generates a file containing all extracted facts, leveraging extraction algorithms similar to those proposed in prior work \cite{QuickInsights, Top-kInsights} (\textbf{R1}).
% In the second stage, the system groups the extracted facts, constructs intra-group similarity matrices, and computes outlier scores (\textbf{R2, R3}), thereby preparing the data for subsequent visualization and exploration.
The Upload Panel consists of two stages. In the first stage, users upload a dataset and configure the subspace and measure attributes. Ordinal subspace attributes can represent sequential variables, such as rankings or years, while an aggregation method can be assigned to each measure attribute. Once the configuration is confirmed, the system automatically extracts the eligible data facts using the procedure described in \hyperref[sec:facts_extraction]{Sec.~\ref*{sec:facts_extraction}} (\textbf{R1}). In the second stage, the system groups the extracted facts, constructs intra-group similarity matrices, and computes their outlier scores (\textbf{R2}, \textbf{R3}), thereby preparing them for visualization and exploration.

\textbf{Overview Panel.}
After fact processing is completed, the Overview Panel presents a matrix-based overview of all fact groups, as shown in \hyperref[fig:VisualExplanation]{Fig.~\ref*{fig:VisualExplanation}A} (\textbf{R4}). Because each group is defined by a unique combination of breakdown and measure attributes, we adopt a matrix layout in which the top row encodes breakdown attributes (\hyperref[fig:VisualExplanation]{Fig.~\ref*{fig:VisualExplanation}A1}) and the left column encodes measure attributes (\hyperref[fig:VisualExplanation]{Fig.~\ref*{fig:VisualExplanation}A2}). Each matrix cell therefore corresponds to a distinct fact group defined by a particular combination of breakdown and measure attributes. Different fact types are represented using distinct colors, with the legend displayed at the bottom of the panel (\hyperref[fig:VisualExplanation]{Fig.~\ref*{fig:VisualExplanation}A3}).
Within each cell, the top progress bar indicates the proportion of facts in that group relative to the full set of extracted facts (\hyperref[fig:VisualExplanation]{Fig.~\ref*{fig:VisualExplanation}A4}). 
The lower portion of each cell displays a scatterplot preview of the contained facts, positioned according to pairwise similarity through dimensionality reduction. 
\revision{Different methods emphasize different structural properties: t-SNE and UMAP preserve local neighborhood relationships \cite{nonato2018multidimensional}, PCA prioritizes variance \cite{DimensionReduction2}, and MDS preserves pairwise dissimilarities \cite{borg2005modern}. The method can be selected according to the analytical objective; in our implementation, we use MDS to better preserve similarities among facts.}
% The lower portion of the cell displays a preview scatterplot of the contained facts, positioned according to pairwise similarity using Multi-Dimensional Scaling (MDS). 
% \revision{Unlike non-linear techniques such as t-SNE or UMAP, which prioritize local neighborhood structures at the expense of global distance fidelity \cite{nonato2018multidimensional}, or linear methods like PCA that prioritize variance over complex non-linear relationships \cite{DimensionReduction1}, MDS directly preserves global pairwise dissimilarities \cite{borg2005modern, DimensionReduction2}, thereby ensuring that the visual distance between clusters faithfully reflects their structural divergence, which facilitates both intuitive and accurate outlier detection among data facts.}
Facts are further organized by their focus, such that facts sharing the same focus are grouped into a single cluster. To visually delineate clusters, we render a smoothed convex hull around each cluster: facts belonging to the same cluster are enclosed by an outer polygon, which is padded, clipped to the canvas, and converted into a smooth closed path drawn as a dashed outline in the cluster color.
Cells encode outlier information using distinct visual styles. Outlier groups, as illustrated in \hyperref[fig:VisualExplanation]{Fig.~\ref*{fig:VisualExplanation}A4}, are highlighted with a light red background and displayed with a slightly larger cell size. In addition, a donut chart appears in the top-right corner to indicate the proportion of outlier facts within that group. Individual outlier facts within the scatterplot are emphasized using a red halo.
Non-outlier groups, shown in \hyperref[fig:VisualExplanation]{Fig.~\ref*{fig:VisualExplanation}A5}, are displayed with a light blue background and a smaller cell, and the outlier proportion indicator is omitted for brevity. Each cell is interactive, and selecting a cell reveals its detailed information in the Main Panel.

\begin{figure*}[t]
  \centering
  \includegraphics[width=1\textwidth]{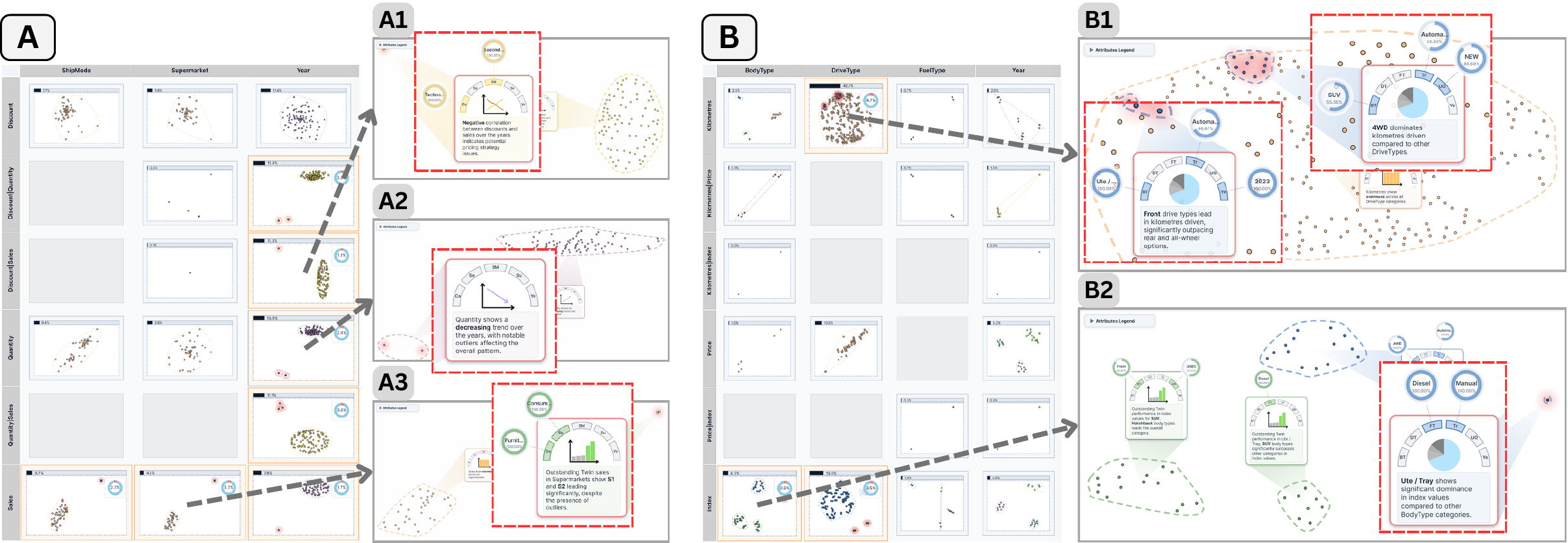}
  \caption{Usage Scenarios. This figure demonstrates the Overview Panel and the corresponding Cluster Map view in the Main Panel after selecting a cell. Panel (A) corresponds to Usage Scenario 1, while panel (B) corresponds to Usage Scenario 2. In each expanded Cluster Map view, the labels of outlier clusters are highlighted with red rectangles.}
  \label{fig:usagescenario}
\end{figure*}

\textbf{Main Panel.}
The Main Panel is organized into four views: (1) Cluster Map, (2) Fact Detail, (3) Fact Flow, and (4) Cluster Insights, as shown in \hyperref[fig:VisualExplanation]{Fig.~\ref*{fig:VisualExplanation}B}.

\textit{1) Cluster Map.}
This view is an enlarged version of the selected cell from the Overview Panel and enables a detailed exploration of data fact clusters organized by focus, as shown in \hyperref[fig:VisualExplanation]{Fig.~\ref*{fig:VisualExplanation}B1} (\textbf{R5}). Within this view, facts belonging to the same focus are grouped into clusters and enclosed by smooth hull boundaries to visually distinguish cluster structures. Clusters containing outlier facts are highlighted differently from those containing only non-outlier facts, allowing analysts to quickly identify unusual patterns. For example, \hyperref[fig:VisualExplanation]{Fig.~\ref*{fig:VisualExplanation}B2} illustrates a typical non-outlier cluster, while \hyperref[fig:VisualExplanation]{Fig.~\ref*{fig:VisualExplanation}B3} shows an outlier cluster that contains facts with high outlier scores, each highlighted with a red halo.
A key addition in this view is the cluster label, a rectangular box surrounded by donut charts. It summarizes important information about each cluster. The label consists of three components. The first component is an icon that conveys the dominant characteristics of the cluster, determined by fact type and focus. The second component is a brief natural-language description generated with LLM assistance, based on the group’s breakdown and measure as well as the cluster’s type and focus, as illustrated in \hyperref[fig:VisualExplanation]{Fig.~\ref*{fig:VisualExplanation}B6} (\textbf{R6}). The third component summarizes the cluster’s contextual subspace analysis. In this design, subspace attributes are arranged along a radial attribute band (\hyperref[fig:VisualExplanation]{Fig.~\ref*{fig:VisualExplanation}B5}), where each band segment corresponds to one attribute and is displayed in abbreviated form, with the attribute legend shown in the upper-left corner of the view (\hyperref[fig:teaser]{Fig.~\ref*{fig:teaser}C1}).
Because each subspace consists of a set of attribute–value filters, we compute each filter’s support within a cluster, namely the proportion of facts containing that filter, to identify frequently occurring subspace filters. When the support of a filter exceeds 50\%, it is visualized as a donut chart connected to the corresponding attribute segment, where the slice encodes the dominant attribute value and its frequency (\hyperref[fig:VisualExplanation]{Fig.~\ref*{fig:VisualExplanation}B4}). This design enables analysts to quickly identify salient contextual subspace filters without manually inspecting individual facts. These dominant filters reveal the common contexts in which facts of particular types or focuses tend to occur, thereby providing additional analytical insight.
By default, labels are placed outside the cluster hull. We sample candidate positions along the hull perimeter, select a nonoverlapping position greedily, and then apply a force-directed adjustment to reduce remaining collisions. In addition, users can manually reposition labels via direct manipulation to suit their preferences. To mitigate occlusion, labels become semi-transparent on hover, allowing users to inspect the underlying facts without obstruction.

\textit{2) Fact Detail.}
When a fact is selected, this view presents its chart, score, subspace, and outlier score in detail. The left side shows a visual chart of the selected data fact, with the chart format determined by the fact type shown in the title. The right side displays the associated metadata, including the score, outlier score, and subspace. Subspace attributes are represented as cards, and non-wildcard attributes are highlighted to indicate the active contextual constraints. An example of this encoding is shown in \hyperref[fig:VisualExplanation]{Fig.~\ref*{fig:VisualExplanation}B7}, where the highlighted pills represent the attribute–value filters that define the subspace. In this example, the subspace is constrained only by the filter $\{FuelType = Diesel\}$.

\textit{3) Fact Flow.}
This view provides a visual overview of how data facts in the selected group are distributed across several analytical dimensions using a Sankey-style diagram \cite{SankeyDiagram}. The flow begins with the outlier status in the first column, distinguishing between outlier and non-outlier facts. These facts then flow into the second column, which represents the fact type, and finally into the third column, which represents the focus of each fact. The width of each flow encodes the number of facts belonging to that category, allowing analysts to quickly understand how facts are distributed and connected across outlier status, type, and focus. Additionally, hovering over a flow segment reveals the exact number of facts represented. By visually tracing these flows, analysts can easily observe the composition and relationships among different categories in the selected group, as illustrated in \hyperref[fig:VisualExplanation]{Fig.~\ref*{fig:VisualExplanation}B8}.

\textit{4) Cluster Insights.}
\revision{This view delivers a coherent, natural-language explanation of each cluster, allowing users to select clusters of interest from an interactive bar and view details below, thereby reducing cognitive load and supporting faster understanding (\textbf{R6}). Drawing on established prompt-engineering practices \cite{PromptEngineering1, PromptEngineering2}, we design a staged, role-based prompting pipeline implemented via the OpenAI API using \texttt{gpt-5.4-mini} with the temperature set to 0.2. Inspired by recent advances in task decomposition and structured prompt engineering \cite{ChainOfThought, LeastToMost, StructuredPrompting}, the pipeline decomposes the summarization task into three sequential stages rather than relying on a single monolithic prompt, helping mitigate hallucinations and reduce cascading reasoning errors. First, a \textit{meaning analyzer} translates abstract schema attributes (i.e., breakdowns, measures, types, and focuses) into plain-language terminology. Second, a \textit{subspace analyzer} contextualizes the cluster by evaluating its frequently occurring subspace filters and the subspaces of data fact outliers. Third, a \textit{group summarizer} ingests these intermediate analytical outputs to synthesize a cohesive, cluster-level narrative.}
An example of the generated summary is shown in \hyperref[fig:VisualExplanation]{Fig.~\ref*{fig:VisualExplanation}B9}.

To facilitate initial use, we incorporate an onboarding feature that automatically guides analysts through the system’s concepts, functionalities, and visual encodings whenever they enter a new interface state.

\section{Usage Scenarios}
\label{sec:scenario}
% To further demonstrate the usefulness and effectiveness of our system, 
We present two usage scenarios using public datasets to illustrate how analysts can explore data and analyze data fact outliers with the system.

\subsection{Scenario 1: Supermarket Sales}
\label{subsec:scenario1}
In the first scenario, we use a supermarket sales dataset\footnote{\url{https://www.kaggle.com/datasets/yikai6/supermarket-resale}} to demonstrate how \sysname assists users in analyzing data fact outliers during exploratory analysis. Before using the system, the analyst has already conducted a rough inspection of the dataset but has not identified any particularly surprising findings. After loading the dataset, extracting data facts, and performing the subsequent analysis, the system identifies 14 cells containing facts, among which seven groups include outlier facts that deserve further inspection, as shown in \hyperref[fig:usagescenario]{Fig.~\ref*{fig:usagescenario}A}.

The analyst first explores the relationship between discounts and sales over time. After opening the corresponding cell, as shown in \hyperref[fig:usagescenario]{Fig.~\ref*{fig:usagescenario}A1}, the analyst finds two clusters of facts. The larger cluster indicates a positive correlation between discounts and sales, which is consistent with common expectations that higher discounts tend to stimulate purchases. However, the analyst is surprised to discover another cluster showing a negative correlation between discounts and sales. By examining the associated subspaces, the analyst finds that this pattern occurs for technology products shipped by second-class delivery. This suggests that higher discounts do not always increase sales and may even reduce them in some contexts.

Next, the analyst investigates which supermarkets dominate sales. After selecting the relevant cell, as shown in \hyperref[fig:usagescenario]{Fig.~\ref*{fig:usagescenario}A2}, the analyst finds that, in most cases, sales are distributed relatively evenly across the six supermarkets. However, one outlier cluster reveals a different pattern: for furniture products purchased by consumer customers, supermarkets S1 and S2 together account for a much larger share of sales than the other four supermarkets. This finding suggests that S1 and S2 may be particularly competitive in the consumer furniture market, possibly because they place greater emphasis on this category or employ more effective sales strategies.

The analyst then turns to quantity trends over the years. Most facts in this group show an increasing trend, while only two facts exhibit a decreasing trend, as shown in \hyperref[fig:usagescenario]{Fig.~\ref*{fig:usagescenario}A3}. Since the automatically generated labels do not reveal any obvious common context for these decreasing-trend facts, the analyst inspects them individually. This closer examination shows that one decreasing trend appears among home office orders delivered by first-class shipping, while another appears for corporate customers purchasing office supplies. These findings help the analyst identify specific contexts in which quantity changes differ from the overall upward trend.

\subsection{Scenario 2: Australian Car Sales}
\label{subsec:scenario2}
In the second scenario, the analyst uses our system to explore an Australian car sales dataset\footnote{\url{https://www.kaggle.com/datasets/nelgiriyewithana/australian-vehicle-prices}}. Compared with the first scenario, this dataset is more complex, making it more difficult to identify data fact outliers through straightforward inspection. After automatic fact extraction and subsequent processing, the system identifies 18 groups of data facts, among which three groups contain fact outliers that merit closer examination, as shown in \hyperref[fig:usagescenario]{Fig.~\ref*{fig:usagescenario}B}.

The analyst first explores the group describing the relationship between kilometers driven and drive type. After opening this group, as shown in \hyperref[fig:usagescenario]{Fig.~\ref*{fig:usagescenario}B1}, the distribution of data facts appears relatively complex at first glance. To gain an initial understanding, the analyst refers to the Fact Flow view, which provides a clear visual overview of how data facts are distributed across categories.
From this overview, the analyst observes that, in most contexts, kilometers driven are relatively evenly distributed across different drive types. However, two outlier clusters exhibit distinctly different patterns. In one cluster, the driving distance is dominated by front-wheel-drive vehicles, whereas in the other it is dominated by four-wheel-drive vehicles.
By inspecting the labels and accompanying summaries, the analyst finds that one cluster is primarily associated with new vehicles, automatic transmission, and the SUV category. Within this cluster, four-wheel-drive vehicles account for a large share of kilometers driven, suggesting a strong association between vehicle type, drivetrain, and usage patterns. In the other cluster, front-wheel-drive vehicles dominate the driving distance and are frequently associated with utility vehicles from 2023 with automatic transmission. The summary panel further helps the analyst confirm and interpret these patterns.

The analyst then examines another group of interest: the relationship between the number of vehicles sold and body type. Using count aggregation, this group can be interpreted as reflecting sales volume. After selecting the corresponding cell, as shown in \hyperref[fig:usagescenario]{Fig.~\ref*{fig:usagescenario}B2}, the analyst inspects the Fact Flow view to understand the distribution of data facts. The visualization reveals two types of data facts and four distinct focuses, each corresponding to a cluster. Compared with the previous group, this one exhibits a more balanced distribution of \textit{Dominance} and \textit{Outstanding Top 2} facts.
A closer inspection of the \textit{Dominance} facts shows that most are associated with SUV sales dominance, while only a single fact indicates dominance by utility vehicles. This observation motivates further exploration of the clusters. The analyst finds that many SUV-dominant facts are associated with contexts involving all-wheel-drive vehicles and automatic transmission, suggesting a strong co-occurrence between these attributes and higher SUV sales. In contrast, the utility vehicle dominance appears in a more specific context, namely diesel vehicles with manual transmission, and is accompanied by a relatively high outlier score of 0.67. Overall, these findings suggest that SUV sales are frequently associated with a range of common configurations, whereas utility vehicles tend to become dominant only under more specific conditions. By combining these observations with information from the Cluster Insights view, the analyst gains a comprehensive understanding of the characteristics, underlying patterns, and contextual associations within each cluster. 
These findings clarify how vehicle characteristics relate to driving patterns and sales outcomes.
% Together with earlier findings on kilometers driven, this analysis provides a more complete picture of how vehicle characteristics relate to both driving patterns and sales outcomes.

\section{User Interview}
To evaluate the usability and effectiveness of \sysname, we conducted in-depth interviews with 12 participants. In this section, we describe the participants, study procedure, and results derived from interviews.

\subsection{Participants}
We recruited 12 participants (P1–P12) from a university population (3 female, 9 male; $age_{mean} = 23.92$, $age_{sd} = 2.78$). All reported normal or corrected-to-normal vision, no color-vision deficiencies, relevant academic backgrounds, and at least one year of experience with data analysis and visualization tools.
\attention{The study was approved by our university's Institutional Review Board (IRB), and informed consent was obtained from all participants.}

\subsection{Procedure}
The interview consisted of four stages and lasted approximately one hour. In the first stage, we provided a brief introduction to the system, including its background and key concepts (e.g., data facts and outlier scores), to familiarize participants with its purpose. In the second stage, we used the dataset described in \hyperref[subsec:scenario1]{Sec.~\ref*{subsec:scenario1}} to demonstrate the system’s functionality and train participants on how to use and explore it. Participants were encouraged to ask questions during this stage.
In the third stage, participants conducted a task-based, think-aloud exploration using the dataset from \hyperref[subsec:scenario2]{Sec.~\ref*{subsec:scenario2}}, 
\revision{where they were asked to identify an outlier group using the Overview Panel, select a data fact outlier within it, interpret the meaning of a specific cluster, and use the Main Panel views to further explore data fact outliers and enhance their understanding.}
Afterward, they were invited to freely explore the system and provide feedback, suggestions, and insights.
In the final stage, participants completed a post-study questionnaire consisting of 11 questions. Nine closed-ended questions used a 5-point Likert scale \cite{LikertScale} (from Strongly Disagree to Strongly Agree) to evaluate different aspects of the system, including the onboarding feature (Q1), the Overview Panel (Q2–Q3), the Main Panel (Q4–Q7), and overall usability (Q8–Q9). Two open-ended questions (Q10–Q11) were included to collect additional feedback and suggestions for improvement.
All participants provided informed consent prior to the study and received compensation of SGD~15 upon completion.

\begin{figure*}[t]
  \centering
  \includegraphics[width=0.95\textwidth]{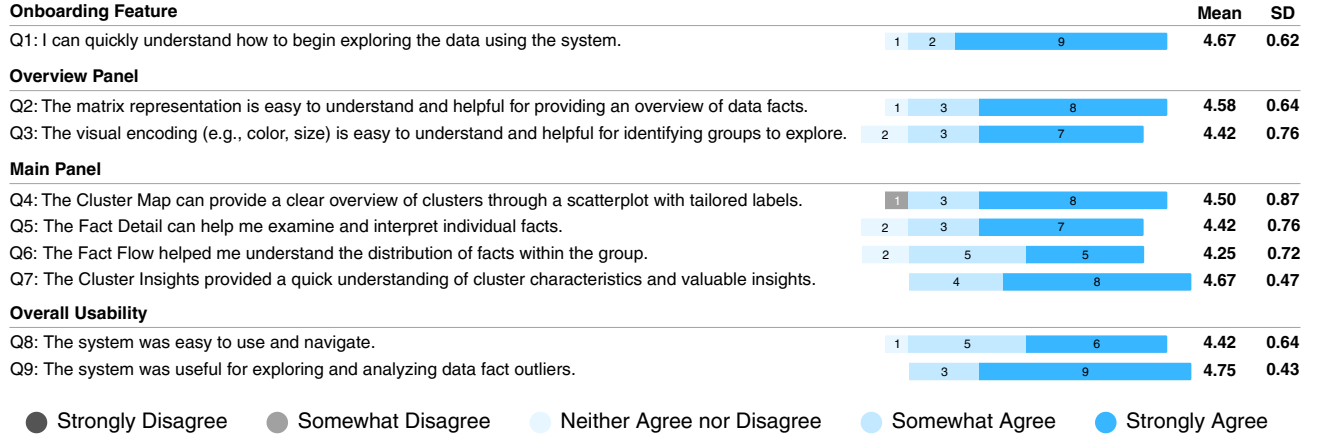}
  \caption{User interview questionnaire results. The results are organized into four parts: Onboarding (Q1), Overview Panel (Q2--Q3), Main Panel (Q4--Q7), and Overall Usability (Q8--Q9). The figure presents the closed-ended questions, their corresponding components, and the mean and standard deviation for each question.}
  \label{fig:interview}
\end{figure*}

\subsection{Results}
\hyperref[fig:interview]{Fig.~\ref*{fig:interview}} presents participants’ responses to the closed-ended questions (Q1–Q9). Overall, the system received positive feedback from participants. In the following, we provide a detailed analysis of user feedback based on these results.

\textbf{Onboarding Feature.} Overall, the onboarding feature received a high rating (Q1: $M = 4.67$). Participants P1 and P5 noted that the onboarding process was particularly helpful for first-time use. However, P2 found the onboarding tour helpful but suggested that certain aspects could be explained in greater detail, as understanding several key concepts is essential for effectively using the system. Consequently, P2 assigned a rating of 3 for this question.

\textbf{Overview Panel.} \revision{Participants generally assigned high ratings to the Overview Panel}. \attention{The matrix representation was considered effective in providing an overview of the distribution of data facts (Q2: $M = 4.58$), while participants found the visual encoding helpful for identifying groups to explore (Q3: $M = 4.42$)}. During task-based exploration, most participants first located cells with a red background and then examined the corresponding distributions in detail. In contrast, P4 and P5 relied more on percentage values to select cells, tending to prioritize those containing more data facts; they described the encoding as both concise and comprehensive.
However, some participants identified limitations. P11 noted that while the matrix representation offers a clear overview of grouped data, it lacks a global view of all data facts within a single unified visualization. Additionally, P6 found the visual encoding of cells insufficiently salient, initially focusing only on the distribution plots below. P6 suggested that key indicators, such as size and color used to represent outlier groups, should be more visually prominent and easier to compare across cells.

\textbf{Main Panel.} \revision{Overall, participants provided positive feedback across these components}: \attention{the Cluster Map view effectively conveys the distribution of data facts within clusters (Q4: $M = 4.50$), the Fact Detail view supports clear inspection of individual facts (Q5: $M = 4.42$), and the Fact Flow view offers an intuitive visual summary (Q6: $M = 4.25$). The Cluster Insights view provides valuable textual explanations (Q7: $M = 4.67$).}
Despite the generally positive feedback, participants identified several areas for improvement. For the Cluster Map view, P7 suggested removing the breakdown attribute from the label ring bands, as it is not filterable under the current constraints and therefore unnecessary to include. P10 commented that the label text is too small and expressed less interest in distributional views, preferring concise, high-level insights.
Regarding the Fact Detail view, P9 noted that although the feature is useful, it may be redundant in practice, as facts within the same cluster often exhibit similar patterns; thus, the Cluster Map view alone may suffice for analysis.
For the Fact Flow view, P2 found it helpful but suggested that it may be unnecessary when the distribution within a group is relatively simple. Similarly, P11 observed that it serves a role similar to the Cluster Map view and suggested aligning their layouts to reflect their comparable overview functions.
For the Cluster Insights view, we observed differing user preferences. Participants P3, P4, and P8 found the LLM-assisted summaries highly beneficial, as they reduce cognitive effort and provide intuitive, text-based insights; P8 further suggested elevating this component to a primary view to better support exploration. In contrast, P9 preferred visual representations over textual descriptions and reported a reluctance to read longer textual summaries during analysis.

\textbf{Overall Usability.} Overall, participants found the system easy to use and navigate (Q8: $M = 4.42$) and effective for exploring and analyzing data fact outliers (Q9: $M = 4.75$). P1 noted that the concept of data fact outliers was novel and enabled exploration across different groups in ways she had not previously considered. P11, who works as an analyst, expressed interest in gaining a deeper understanding of the system and potentially applying it in future practice. 
\revision{P9, who has a strong background in visualization, noted that the Overview and Main Panels support both global overview and drill-down analysis, offering a clearer design than combining all information within a single view.}
However, she also emphasized that a solid understanding of the fundamental concepts is important for effectively using the system in analysis. 

\textbf{Open-ended Questions.} In this section, we evaluated participant preferences regarding component utility (Q10) and gathered suggestions for improvement (Q11). Regarding utility, participants expressed two main preferences: P1, P2, P6, P7, P9, and P11 favored the Cluster Map view for its informative labels and effective distribution overview, while the remaining participants preferred the Cluster Insights view for providing intuitive insights that minimize cognitive load and learning effort.
Participants also provided several suggestions for improvement. For the interface, P5 proposed a color configuration panel to customize data fact encodings, while P11 and P12 suggested increasing the prominence of the Fact Flow via layout adjustments or richer visual encodings. Regarding the LLM features, P8 suggested expanding the LLM-based summary to a global view with natural-language exploration, whereas P9 advised reducing summary verbosity in favor of more intuitive visualization.

% \textbf{Open-ended Questions.} In this section, we asked participants which component they found most useful (Q10) and invited suggestions for improvement (Q11). The responses revealed two primary preferences. Participants P1, P2, P6, P7, P9, and P11 identified the Cluster Map as the most helpful component, citing its ability to convey the overall distribution of groups alongside informative labels. In contrast, the remaining participants favored the Cluster Insights, noting that it provides intuitive insights while reducing cognitive and learning effort.
% Participants also provided several suggestions for improvement. P5 highlighted limitations in the current color design, suggesting that different types of data facts could be distinguished using customizable color encodings, potentially supported by an additional color configuration panel. P8 proposed extending the LLM-assisted summary to a global view, enabling users to interact with the system using natural language to guide exploration. In contrast, P9 suggested reducing the verbosity of the LLM-generated summaries and incorporating more visual representations to support intuitive understanding.
% Additionally, P11 and P12 emphasized the importance of the Fact Flow as a visual overview and suggested enhancing its prominence through layout adjustments or richer visual encodings.

\section{Discussion and Future Work}

In this section, we discuss limitations of our system and outline possible directions for future work.

\textbf{Extending Supported Data Fact Types.}
Currently, our system supports eight predefined data fact types, which may not capture all analytical patterns of interest, such as seasonality \cite{QuickInsights}. To address this, future work could explore mechanisms for user-defined pattern specification, allowing analysts to describe new fact types that can be integrated into the extraction algorithms for more flexible analysis.

\textbf{Supporting Diverse User Preferences.}
Our interviews revealed that users exhibit differing preferences in how they interact with the system. For example, P8 preferred the LLM-assisted summary and suggested expanding its role, whereas P9 favored visualization-based exploration and preferred minimizing textual descriptions. To accommodate such differences, future work could introduce adaptive or customizable interfaces, such as a toggle mechanism that allows users to switch between text-driven and visualization-driven exploration modes.

\revision{
\textbf{Supporting Natural-Language Interaction.} While existing LLM-based visual analytics tools are effective at generating general explanations \cite{InsightPilot, InsightsSurvey}, \sysname focuses on targeted data fact outlier analysis by using deterministic scoring to identify data fact outliers. However, the current LLM pipeline only generates static summaries for detected clusters and does not allow users to further explore the results. Future work could integrate an LLM-based conversational interface with natural-language query capabilities, enabling users to ask follow-up questions, request clarification, and investigate specific patterns or data fact outliers in greater depth.
}

% \textbf{Advancing beyond One-Way LLM Generation.} While existing LLM-based visual analytics tools are effective at generating general explanations \cite{InsightPilot, InsightsSurvey}, \sysname is optimized for targeted outlier analysis by using deterministic scoring to identify data fact outliers. However, its current LLM pipeline remains one-way: it generates static summaries for detected clusters but does not support user interaction or follow-up questions. Future work could extend this pipeline into a bidirectional conversational interface, enabling users to ask questions about unclear patterns, seek further clarification, and interactively refine their understanding of data fact outliers.

\revision{
\textbf{Enriching Outlier Semantics and Relationships.} Currently, our system uses a binary classification of data fact outliers and focuses on intra-group exploration, overlooking the distinct semantic roles that outliers may play. For example, an increasing trend among predominantly decreasing trends contradicts the prevailing pattern, whereas a \textit{Dominance} fact focused on \textit{Country B} among facts predominantly focused on \textit{Country A} represents a distinctive alternative focus. Moreover, the relationships between group-level outliers and global dataset patterns remain unexplored, and the system does not identify which breakdown or measure attributes are associated with the highest proportions of outliers. Future work will address these limitations by enriching data fact outlier semantics and linking intra-group findings to the global matrix to support broader analysis.
}

% \textbf{Enriching Outlier Semantics and Relationships.} Currently, our system uses a binary classification of outliers and focuses on intra-group exploration. However, outliers may have different semantic roles: some conflict with broader patterns, whereas others represent unique but complementary cases. Furthermore, the relationships between group-level outliers and global dataset patterns remain unexplored. The system also does not identify which breakdown or measure attributes are associated with the highest proportions of outliers. Future work will address these limitations by enriching outlier semantics and linking intra-group findings to the global matrix to support broader analysis.

\textbf{Improving Scalability.}
When applied to large datasets or high volumes of data facts, the current extraction algorithm and Overview Panel face performance and visualization scalability limits. Future work will focus on optimizing large-scale fact extraction and introducing hierarchical or sampling-based techniques to manage complexity.

\section{Conclusion}
We present \sysname, a visual analytics system for exploring data fact outliers. The system groups data facts into consistent analytical scopes and introduces a unified scoring framework that combines distribution-based and pattern-based scores. We design an interactive interface with coordinated visualizations and LLM-assisted summaries to help analysts examine outlier clusters and interpret individual facts. Through two usage scenarios and interviews with 12 participants, we demonstrate that \sysname\ enables effective exploration of data fact outliers.
Future work will support user-defined fact types, customizable interfaces, natural-language interaction, richer semantics, and greater scalability.

% In the future, we plan to extend the framework with more advanced scoring strategies and improved scalability for larger and more complex datasets.
% \wy{Add one more future direction here.}

%% if specified like this the section will be omitted in review mode
\acknowledgments{%
This project is supported by the Ministry of Education, Singapore, under its Academic Research Fund Tier 1 (NTU Tier 1, RG104/25) and the NTU Start Up Grant awarded to Yong Wang. Any opinions, findings, conclusions, or recommendations expressed in this material are those of the author(s) and do not reflect the views of the Ministry of Education, Singapore.
}

\bibliographystyle{abbrv-doi-hyperref}

\bibliography{bibliography}

\appendix % You can use the `hideappendix` class option to skip everything after \appendix
\crefalias{section}{appendix} % this is to make sure that cleverref switches to referring to Appx. X from here on

\end{document}